\documentclass[paper,empty]{ieice}
\usepackage[dvipdfm]{graphicx}
\usepackage{color}
\usepackage{relsize}
\usepackage{algorithm}
\usepackage{algorithmic}

\renewcommand{\algorithmicrequire}{\textbf{Input:}}
\renewcommand{\algorithmicensure}{\textbf{Output:}}
\usepackage{graphicx}
\usepackage{latexsym}
\usepackage{amsmath}
\usepackage{amssymb}
\usepackage{bm}
\usepackage{amsmath,amssymb}
\usepackage{graphicx}

\newcommand{\sgn}{\mathrm{sgn}}
\newcommand{\tentative}{\color{black}\noindent{{{\tt TENTATIVE DECISION}}} \color{black}}
\newcommand{\ctov}{\color{black}\noindent{{{\tt CHECK TO VARIABLE}}} \color{black}}
\newcommand{\vtoc}{\color{black}\noindent{{{\tt VARIABLE TO CHECK}}} \color{black}}
\newcommand{\initialization}{\color{black}{\noindent{{{\tt INITIALIZATION}}} }\color{black}}

\newcommand{\GF}{\mathrm{GF}}

\vol{93}
\no{11}
\field{A}
\SpecialSection{Information Theory and Its Applications}
\title{Fourier Domain Decoding Algorithm of Non-Binary LDPC codes for Parallel Implementation}
\authorlist{
 \authorentry[kenta@comm.ss.titech.ac.jp]{Kenta KASAI}{m}{TITECH}
 \authorentry[sakaniwa@ss.titech.ac.jp]{Kohichi SAKANIWA}{f}{TITECH}
}
\affiliate[TITECH]{Dept.\ of Communications and Integrated Systems, 
Tokyo Institute of Technology
2-12-1, O-Okayama, Meguro-ku, Tokyo, 152-8550, Japan}

\received{2010}{2}{19}
\revised{2010}{5}{24}

\begin{document}
\maketitle
\begin{summary}
For decoding non-binary low-density parity-check (LDPC) codes, 
logarithm-domain sum-product (Log-SP) algorithms were proposed for reducing quantization effects of 
SP algorithm in conjunction with FFT. 
Since FFT is not applicable in the logarithm domain, the computations required at check nodes in the Log-SP algorithms are 
computationally intensive. 
What is worth, check nodes usually have higher degree than variable nodes. 
As a result, most of the time for decoding is used for check node computations, 
which leads to a bottleneck effect. 
In this paper, we propose a Log-SP algorithm in the Fourier domain. 
With this algorithm, the role of variable nodes and check nodes are switched. 
The intensive computations are spread over lower-degree variable nodes, 
which can be efficiently calculated in parallel.
Furthermore, we develop a fast calculation method for the estimated bits and syndromes in the Fourier domain. 
\end{summary}

\section{Introduction}
In 1963, Gallager invented binary low-density parity-check (LDPC) codes \cite{gallager_LDPC}.
Due to the sparseness of the code representation, 
LDPC codes are efficiently decoded by sum-product decoders (SP) \cite{910572} or Log-SP decoders \cite{910577}. 
The Log-SP is also known as belief propagation.
By the powerful method {\it density evolution} \cite{910577}, invented by Richardson and Urbanke,
messages of Log-SP decoding are statistically evaluated. 
The optimized LDPC codes can realize the reliable transmissions at rate  very close to the Shannon limit \cite{richardson01design}.

The binary LDPC codes are defined by sparse parity-check matrices over  $\GF(2)$. 
On the other hand, the non-binary LDPC codes are defined by sparse parity-check matrices over  $\GF(2^m)$ for $2^m>2$. 
Non-binary LDPC codes are invented by Gallager \cite{gallager_LDPC}  and, Davey and MacKay \cite{DaveyMacKayGFq} found non-binary LDPC codes can outperform binary ones. 
Non-binary LDPC codes have captured much attention recently due to their decoding performance \cite{4717467,1512327,5089504,5191110,4939224}. 

It is known that irregularity of Tanner graphs help improve the decoding performance of binary LDPC codes \cite{richardson01design}. 
On the other hand, it is not the case for the non-binary LDPC codes. 
The $(j=2, k)$-regular non-binary LDPC codes over $\GF(2^m)$ are empirically known \cite{4641893} as the best performing codes for $2^m\ge 64$, especially for short code length. 
This means that, for designing non-binary LDPC codes, one does not need to optimize degree distributions of Tanner graphs, since $(j=2, k)$-regular non-binary LDPC codes are best. Therefore, we assume $j=2$. 
Furthermore, the sparsity of $(j=2, k)$-regular Tanner graph leads to efficient decoding. 
The coding rate is given by $R=1-2/k$. It can be seen that $k$ gets higher as $R$ tends to $1$. 

In this paper, we deal with the decoding algorithms of non-binary LDPC codes \cite{DaveyMacKayGFq},  which is applicable to the binary LDPC code.
Despite of the efficient and parallel implementation of the Log-SP algorithm, the application of LDPC codes to the industry 
is limited so far. 
This is due to the requirement of large memory devices and computationally intensive non-linear check-node 
computation for the decoders. 
The conventional decoding algorithm for non-binary LDPC codes, compared to the binary counterpart,  
is computationally complex and require more memories to store messages.

Immediate use of the SP algorithm for non-binary LDPC codes over $\GF(2^{m})$ 
requires $O((k-1)2^{2m})$ additions and multiplications per check node and $O(2^m)$ multiplications per variable node. 
By FFT and IFFT \cite{davey_phd_thesis}, 
the check node computation in the SP algorithm is largely reduced 
to $O(km2^m)$ additions and multiplications. 
However, the SP algorithm with FFT is not robust to a quantization effect since 
the messages are recursively multiplied among them. 
The quantization effect can not be observed in normal PC-like computers equipped with 
32-bit FPUs and large memory devices. 
However, the use of such high quantization-level processors and large memory devices prevents realizing
high-throughput decoders. 

In order to avoid the quantization effect or large memory devices, and to realize the high-throughput decoders, the SP algorithms in the logarithm domain \cite{1312606,4155118} have been proposed.
Multiplications are replaced with additions by treating the messages in the logarithm domain, which reduces the quantization effect. 
However, with the logarithm-domain SP algorithm, 
we have to give up the efficient calculation of check node computations, 
since FFT and IFFT can not be applied to the messages in the logarithm domain. 
The check node computation requires at most $O((k-1)2^{2m})$ times of $\ln(\cdot)$ $\exp(\cdot)$ and additions, or
is approximated by simpler calculations \cite{1312606,4155118}, which still requires much higher computations 
than variable nodes. 

In summary, we face the following problems when using the SP algorithm in the logarithm domain. 
\begin{itemize}
 \item FFT and IFFT are not applicable to the check node computations. 
 \item The check node computation requires much higher computation than that of variable nodes.  
What is worse, check nodes have higher degree $k=2/(1-R)>2$ than variable nodes of degree 2. 
\end{itemize}
It can be seen that 
the most intensive computations are calculated at the most crowded nodes, i.e., nodes of the highest degree. 
These problems cause a bottleneck at check node computation. 
Such a  bottleneck problem can not be solved even by using fully node-parallel processors,  since the node computations are triggered by incoming messages. 
Most of the time for decoding is used for the check node computations. 

In this paper, we propose a decoding algorithm whose messages consistently stays in the Log-Fourier domain. 
With this algorithm, the computations of variable nodes and check nodes are switched. 
Check nodes still have higher degree $k$ than variable nodes of degree 2, but the computations 
for the check node become much lower. 
On the other hand, variable nodes require much more computations, but the degree is only 2. 
Consequently, the computations for the decoding are spread all over the nodes. 

Note that our interest is not reducing the total amount of computations for decoding 
but reducing the most intensive computation among all the node. 
Such reduction is important in the situation that the computation for each node is operated in parallel. 
The proposed algorithm removes 
the obstacles which block squeezing the potential of parallel implementation of LDPC codes. 
\section{Conventional Decoding Algorithms and Bottleneck Problem}
\label{051950_19Nov09}

For simplicity, we consider a non-binary $(j, k)$-regular LDPC code over $\GF(2^m)$. 
The extension  to irregular LDPC codes is straightforward.
Let $N$ be the code length in terms of $\GF(2^m)$ symbol,  then the  number of parity-check constraints is given as $M=jN/k$.
Once a primitive element $\alpha$ of $\GF(2^m)$ is fixed, 
each symbol is given $m$-bit representation \cite[pp.~110]{macwilliams77}. 
For example, with a primitive element $\alpha\in\GF(2^3)$ such that $\alpha^3+\alpha+1=0$, each symbol is represented as
$0=(0,0,0)$, $1=(1,0,0)$, $\alpha=(0,1,0)$, $\alpha^2=(0,0,1)$, 
$\alpha^3=(1,1,0)$, $\alpha^4=(0,1,1)$, $\alpha^5=(1,1,1)$ and $\alpha^6=(1,0,1)$. 
We shall interchangeably use the two representations, i.e., $x\in\GF(2^m)$ as a symbol in the Galois field and a $m$-bit vector. 

In the binary representation, the codewords can be viewed as binary sequences of length $mN$. 
Given an $M\times N$ parity-check matrix $H=\{h_{cv}\}$ over $\GF(2^m)$ of column weight $j$ and row weight $k$, 
the non-binary LDPC code defined by $H$ is given as 
\begin{align*}
 &\{x\in\GF(2^m)^N\mid Hx=0\in\GF(2^m)^M\}.
\end{align*}
The $c$-th row of the parity-check matrix represents a parity-check equation 
\begin{align*}
& h_{cv_1}x_{v_1}+\cdots+h_{cv_k}x_{v_k}=0,
\end{align*}
where $h_{cv_i}, x_{v_i}\in \GF(2^m) \text{ for } i=1,\dotsc,k$. 

The Tanner graph of the non-binary LDPC code is given by a bipartite graph of $N$ variable nodes  and $M$ check nodes. 
The $v$-th variable node and $c$-th check node are adjacent each other iff $h_{cv}\neq 0$.
For simplicity, we denote $v$-th variable node and $c$-th check node by $v$ and $c$, respectively. 
Define  $V_c$ as the set of adjacent variable nodes of a check node $c$.  
Similarly,  define  $C_v$ as the set of adjacent check nodes of a variable node $v$.  
For $(j,k)$-regular LDPC codes, we have $|V_c|=k$ and $|C_v|=j$.

The decoding algorithms for binary and non-binary  LDPC are usually viewed as {\itshape message-passing algorithms} 
on the Tanner graphs. 
All the algorithms dealt in this paper involve the following 4 steps. 
\begin{enumerate}
\item \initialization: For each variable node $v (v=1,\dotsc,N)$, the initial message is calculated from the channel output for the $v$-th transmitted symbol. 
The variable node $v$ sends the initial message to the adjacent check node $c$ for $c\in C_v$.
The iteration round $\ell$ is set as $\ell:=0$.
\item \ctov: Each check node $c (c=1,\dotsc,M)$ has $k$ incoming messages sent from its $k$ adjacent variable nodes. The check node $c$ computes the outgoing messages to be sent to the $k$ adjacent variable nodes. 
Increment the iteration round as $\ell:=\ell+1$. 
\item \vtoc: Each variable node $v$ has $j$ incoming messages sent from its $j$ adjacent check nodes. 
With the initial message, the variable node $v$ computes the outgoing messages to be sent to the $j$ adjacent check nodes.
\item \tentative: For each variable node $v$, a tentatively estimated symbol 
$\hat{x}_v^{(\ell)}\in \GF(2^m)$ is calculated from the messages sent from its $j$ adjacent check nodes and the initial message. 
If the tentatively estimated symbols  $(\hat{x}_1^{(\ell)},\dotsc,\hat{x}_N^{(\ell)})$  form a codeword, the decoder outputs the codeword, otherwise repeats the steps 2, 3 and 4.
If the iteration round $\ell$ reaches at a pre-determined number, the decoder outputs {\tt FAIL}.
\end{enumerate}

The subject in this paper is about the decoding algorithms for the non-binary LDPC codes. 
To emphasize the difficulty of problems, we first review the conventional decoding algorithms for non-binary LDPC codes. 
\subsection{Conventional Sum-Product Algorithm for Non-Binary LDPC Code}
\label{053919_6Oct09}
In this section, we review the conventional decoding algorithm \cite{706440},\cite{RaU05/LTHC}
for non-binary LDPC codes over $\GF(2^m)$, i.e.,  the SP algorithm. 
In the SP algorithm, the messages are represented  as probability vectors over $\GF(2^m)$. 
Each message is represented as a vector in $[0,1]^{2^m}$. 
The algorithm is  the symbol-wise maximum a posterior probability (symbol MAP) decoding if the Tanner graph is a tree. 
Even if the Tanner graph is not a tree, due to its sparseness, the algorithm can approximates the symbol-MAP decoding. 
The following describes the conventional SP algorithm \cite{706440},\cite{RaU05/LTHC}. 
\\
\\\initialization: For each variable node $v$, the initial message is given as follows. 
 \begin{align*}
  p_v^{(0)}(x)&= {\Pr(X_v=x|Y_v=y_v)}, \text{ for } x\in \GF(2^m), 
 \end{align*} 
where $X_v$ is the random variable of the $v$-th transmitted symbol, $y_v$ is the channel output of the $v$-th transmitted symbol and $Y_v$ is its random variable. 
The variable node $v$ sends the message $p^{(0)}_{vc}=p^{(0)}_{v}\in [0,1]^{2^m}$ to $c$ for $c\in C_v$.
Set $\ell:=0$.
\\\\\ctov:
Each check node $c$ has $k$ incoming messages $p^{(\ell)}_{vc} \in [0,1]^{2^m} (v\in V_c)$ sent from its $k$ adjacent variable nodes. 
The check node $c$ sends the message $p^{(\ell+1)}_{cv}\in [0,1]^{2^m}$ to $v$ for $v\in V_c$.
\begin{align}
&\tilde{p}^{(\ell)}_{vc}(x) = {p}^{(\ell)}_{vc}(h_{cv}^{-1} x), \text{ for } x\in \GF(2^m)\nonumber\\
&\tilde{p}^{(\ell+1)}_{cv}= \bigotimes_{v'\in V_c\backslash\{v\}}\tilde{p}^{(\ell)}_{v'c},\label{193628_11Feb10} \\
&{p}^{(\ell+1)}_{cv}(x) = \tilde{p}^{(\ell+1)}_{cv}(h_{cv} x),\text{ for } x\in \GF(2^m)\nonumber,
\end{align}
where $p_1\otimes p_2\in [0,1]^{2^m}$ is a convolution of $p_1\in [0,1]^{2^m}$ and $p_2\in [0,1]^{2^m}$. To be precise,  for  $x\in \GF(2^m)$, 
\begin{equation*}
  (p_1\otimes p_2)(x) = \sum_{x_1,x_2\in\GF(2^m):x=x_1+x_2}{p_1(x_1)p_2(x_2)}.
\end{equation*}
We denote $p_1\otimes\cdots\otimes p_k$ by $\bigotimes_{i=1}^{k}p_i$. 
Increment the iteration round as $\ell:=\ell+1$. 
\\\\\vtoc:
Each variable node $v$ has $j$ incoming messages $p^{(\ell)}_{cv}\in [0,1]^{2^m} (c\in C_v)$ sent from its $j$ adjacent check nodes. 
The variable node $v$ sends the message $p^{(\ell)}_{vc}$ to $c$ for $c\in C_v$.
\begin{align*}
p^{(\ell)}_{vc}(x) = p_v^{(0)}(x)\prod_{c'\in C_v\backslash\{c\}}p^{(\ell)}_{c'v}(x) \text{ for } x\in \GF(2^m).
 \end{align*} 
And normalize $p^{(\ell)}_{vc}$ so that $\sum_{x\in\GF(2^m)}p^{(\ell)}_{vc}(x)=1$ as follows. 
\begin{align*}
 p^{(\ell)}_{vc}(x):={p^{(\ell)}_{vc}}(x)/{\sum_{x\in\GF(2^m)}p^{(\ell)}_{vc}(x)}, \text{ for } x\in\GF(2^m). 
\end{align*}
The decoding output does not change even if this normalization step were replaced with
\begin{align*}
 p^{(\ell)}_{vc}(x):={p^{(\ell)}_{vc}}(x)/{p^{(\ell)}_{vc}}(0), \text{ for } x\in\GF(2^m). 
\end{align*}
In this case, the messages are no longer probability vectors. 
\\\\\tentative:
The tentatively estimated symbol $\hat{x}_v^{(\ell)}\in \GF(2^m)$ for the $v$-th transmitted symbol is given as 
\begin{align*}
&\hat{x}_v^{(\ell)}=\mathop{\mathrm{argmax}}_{x\in \GF(2^m)}q_v^{(\ell)}(x),\\
&q_v^{(\ell)}(x):=p_v^{(0)}(x)\prod_{c'\in C_v}p^{(\ell)}_{c'v}(x), \text{ for } x\in \GF(2^m). 
\end{align*} 

The calculation in Eq.~\eqref{193628_11Feb10} is the most complex part of the decoding. 
However, the convolution is efficiently calculated via the Fourier transforms \cite{4155118}. 
For example, the $k$-fold convolution 
\begin{align*}
  q=  \bigotimes_{i=1}^k p_i\in [0,1]^{2^m}
\end{align*}
is  efficiently calculated via the Fourier transform, for $i=1,\dotsc, k$, 
  \begin{align*}
   &P_i(z) := \sum_{x\in\GF(2^m)} p_i(x)(-1)^{z \cdot x}\text{ for }z\in\GF(2^m), 
  \end{align*}
and component-wise multiplications, 
  \begin{align*}
   &Q(z) := \prod_{i=1}^kP_i(z)\text{ for }z\in\GF(2^m), 
  \end{align*}
and the inverse Fourier transform
  \begin{align*}
   &q(x) = \frac{1}{2^m}\sum_{x\in\GF(2^m)} Q(z)(-1)^{z \cdot x}, \text{ for }x\in\GF(2^m), 
  \end{align*}
where $z\cdot x$ is the dot product of the binary representations of $z$ and $x$. 
For example, for 
$z=(1,1,0,1,1,1,0,0)$ and 
$x=(1,0,0,1,1,0,1,1)$, 
$z\cdot x=1+0+0+1+1+0+0+0=3$.
The Fourier transform and inverse Fourier transform are efficiently calculated by FFT and IFFT \cite{davey_phd_thesis} and \cite{4155118}.

By using FFT, the SP algorithm can be viewed as the iteration of FFT and component-wise multiplications. 
Compared to additions, multiplications requires more complex computation devices and higher level quantizations. 
Hence, it is strongly desired to avoid multiplications in decoders, 
in order to meet the demand of the high speed and low quantization level decoders.
\subsection{Logarithm-Domain Sum-Product Decoding for Non-Binary LDPC Codes}
In the SP algorithm, lots of multiplications are needed. 
Transforming the messages to logarithm domain, the multiplications can be done by additions. 
The following algorithm describes the logarithm-domain SP algorithm which is referred to as the Log-SP algorithm \cite{1312606}. 
\\\\\initialization: For each variable node $v$, the initial message is given as follows. 
 \begin{align*}
  \lambda_v^{(0)}(x)&= \ln{(\Pr(X_v=x|Y_v=y_v))}, \text{ for } x\in \GF(2^m)
 \end{align*} 
Each variable node $v$ sends the message $\lambda^{(0)}_{vc}=\lambda^{(0)}_{v}\in[-\infty,0]^{2^m}$ to $c$ for $c\in C_v$.
Set $\ell=0$.
\\\\\ctov:
Each check node $c$ has $k$ incoming messages $\lambda^{(\ell)}_{vc}\in [-\infty,0]^{2^m} (v\in V_c)$ sent from its $k$ adjacent variable nodes. 
The check node $c$ sends the message $\lambda^{(\ell+1)}_{cv}\in [-\infty,0]^{2^m}$ to $v$ for $v\in V_c$.
\begin{align}
&\tilde{\lambda}^{(\ell)}_{vc}(x) = {\lambda}^{(\ell)}_{vc}(h_{cv}^{-1} x), \text{ for } x\in \GF(2^m),\nonumber\\
&\tilde{\lambda}_{cv}^{(\ell+1)}=\boxtimes_{v'\in V_c\backslash\{v\}}\tilde{\lambda}^{(\ell)}_{v'c},\label{195347_11Feb10} \\
&{\lambda}^{(\ell+1)}_{cv}(x) = \tilde{\lambda}^{(\ell+1)}_{cv}(h_{cv} x), \text{ for } x\in \GF(2^m),\nonumber 
 \end{align} 
where $\lambda_1\boxtimes \lambda_2\in [-\infty,0]^{2^m}$ is defined as follows. 
\begin{equation*}
 (\lambda_1\boxtimes \lambda_2)(x) =\ln( \sum_{x_1,x_2\in\GF(2^m):x=x_1+x_2}e^{\lambda_1(x_1)+\lambda_2(x_2)}),
\end{equation*} 
for $x\in\GF(2^m)$. We denote $\lambda_1\boxtimes\cdots\boxtimes \lambda_k$ by $\boxtimes_{i=1}^k\lambda_i$. 
Increment the iteration round as $\ell:=\ell+1$. 
\\\\\vtoc:
Each variable node $v$ has $j$ incoming messages $\lambda^{(\ell)}_{cv} (c\in C_v)\in [-\infty,0]^{2^m}$ sent from its $j$ adjacent check nodes. 
The variable node $v$ sends the message $\lambda^{(\ell)}_{vc}$ to $c$ for $c\in C_v$.
\begin{align*}
{\lambda}^{(\ell)}_{vc}(x)=&\lambda_v^{(0)}(x)+\sum_{c'\in C_v\backslash\{c\}}\lambda^{(\ell)}_{c'v}(x) \text{ for } x\in \GF(2^m). 
 \end{align*} 
And normalize $\lambda^{(\ell)}_{vc}\in [-\infty,0]^{2^m}$ so that $\lambda^{(\ell)}_{vc}(0)=0$ as follows. 
\begin{align*}
 \lambda^{(\ell)}_{vc}(x):={\lambda^{(\ell)}_{vc}}(x)-{\lambda^{(\ell)}_{vc}}(0), \text{ for } x\in\GF(2^m). 
\end{align*}
\tentative:
The tentatively estimated symbol $\hat{x}_v^{(\ell)}\in \GF(2^m)$ for the $v$-th transmitted symbol is given as 
\begin{align*}
&\hat{x}_v^{(\ell)}=\mathop{\mathrm{argmax}}_{x\in \GF(2^m)}\mu_v^{(\ell)}(x),\\
&\mu_v^{(\ell)}(x):=\lambda_v^{(0)}(x)+\sum_{c'\in C_v}\lambda^{(\ell)}_{c'v}(x) \text{ for } x\in \GF(2^m).
\end{align*} 

It can be easily seen that the outputs of this Log-SP algorithm is the same as those of the SP algorithm. 
The check node computation Eq.~\eqref{195347_11Feb10} still is the most complex part of the decoding. 
The computation in Eq.~\eqref{195347_11Feb10} can be viewed as a convolution in the logarithm domain. We refer to this operator $\cdot\boxtimes\cdot$ as {\it the log-convolution}.
Such a log-convolution can not be calculated efficiently by FFT, IFFT and component-wise multiplications, since the messages are transformed in the logarithm domain. 
However, since the Log-SP algorithm does not need multiplications but additions, it is more robust to 
the quantization effects when the messages are stored on a small number of bits \cite{1312606,706440}. 

For computing Eq.~\eqref{195347_11Feb10}, using look-up tables is proposed in \cite{1312606}. 
Declercq et al.~proposed storing only the most contributing messages \cite{4155118}, 
which gave a good trade-off between the decoding complexity and the decoding performance.  
\subsection{Bottleneck Problem}
\label{bottleneck}
Due to the demand of the high speed and low quantization level decoders, 
we can not afford multiplications which are computationally expensive. 
Consequently, one needs to choose the Log-SP algorithm rather than the SP algorithm. 

Compared to the variable node computations, the check node computations have two reasons for being  the bottleneck of the Log-SP algorithm. 
First is obvious as seen so far. 
The computations in variable nodes are simple component-wise additions of message vectors,
while the computations in check nodes need non-linear calculations as in Eq.~\eqref{195347_11Feb10}.

The second reason is that the number of incoming messages sent into check nodes is generally higher than that of variable nodes. 
For $(j,k)$-regular non-binary LDPC codes, variable and check nodes have $j$ and $k$ incoming message vectors, respectively. 
The coding rate $R$ is given as $R=(k-j)/k$. 
Therefore, the number of incoming messages to check nodes are $k/j=1/(1-R)$ times as higher as that of variable nodes. 
The ratio $k/j$ gets higher as $R\to 1$.

Due to the above two asymmetry about computation at variable nodes and check nodes, i.e.,  
the number of incoming messages and the difference of computation functions,  
we face a bottleneck problem of check node computations. 
One may think, in general, bottleneck problems can be solved by using parallel processors to allocate
computation resources intensively to the bottleneck computations. 
However, since the variable and check node computations are triggered by incoming messages, 
the bottleneck problem of check node computations can not be solved even in the situation that fully node-parallel 
processing is possible. 

In the situation that each node-computation is processed in parallel, 
the total decoding time depends on the most complex node-computation among the all nodes. 
In this paper, we propose a decoding algorithm for non-binary LDPC codes,  
which reduces the largest node-computation amount of among all the nodes. 
Note that our interest is not for reducing the total amount of computations for decoding. 
\section{New Fourier and Log-Fourier Sum-Product Algorithms for Non-Binary LDPC Codes}
\label{113552_13Feb10}
In order to reduce the computation amount per check node which is a bottleneck in the Log-SP algorithm,  
we propose a decoding algorithm such that 
the role of variable nodes and check nodes are switched
by initializing  messages by the Fourier transform. 
To be precise, log-convolutions are done at the computation at variable nodes  and 
component-wise additions are done the computation at check nodes . 

As a preparatory algorithm for the Log-Fourier SP algorithm that will be introduced in Section \ref{logsp}, 
firstly, in Section \ref{fsp}, we introduce the SP algorithm in the Fourier domain, which is referred to as the Fourier SP algorithm. 
The messages in the Fourier SP algorithm
are Fourier transformed at the beginning and inverse Fourier transformed at the end. 
The computations at variable node and check nodes, i.e., component-wise multiplications and convolutions,  are switched in the Fourier domain. 
To be  precise, with the proposed algorithm, 
messages are convoluted at variable nodes and  messages are component-wisely multiplied at check nodes. 
The Fourier-SP algorithm is designed so that it outputs  the same decoding results 
as SP and Log-SP algorithms do. 
Thus, the computational intensive tasks, i.e., convolutions are assigned to the variable nodes that have
less incoming messages than check nodes. 
On the other hand, the computationally less intensive tasks, i.e., component-wise multiplications are
assigned to check nodes which have larger incoming messages than variable nodes. 
\subsection{New Fourier Sum-Product Algorithm for Non-Binary LDPC Code}
\label{fsp}
The following describes the Fourier SP algorithm. 
Note again that this is a preparatory algorithm for helping understand the algorithm in the next Section \ref{logsp}. 
\\\\\initialization: 
For each variable node $v$, the initial Fourier-transformed message $P_v^{(0)}\in [-1,1]^{2^m}$ is given as follows. 
 \begin{align*}
  p_v^{(0)}(x)&= \Pr(X_v=x|Y_v=y_v), x\in\GF(2^m)\\
  P_v^{(0)}(z)&= \sum_{x\in\GF(2^m)} {p_v^{(0)}(x)}(-1)^{z\cdot x}, z\in\GF(2^m)
 \end{align*} 
for $z\in \GF(2^m)$. This can be  done via FFT. 
Each variable node $v$ sends the message $P^{(0)}_{vc}=P^{(0)}_{v}\in [-1,1]^{2^m}$ to $c$ for $c\in V_c$.
Set $\ell=0$.
\\\\\ctov:
Each check node $c$ has $k$ incoming messages $P^{(\ell)}_{vc}\in [-1,1]^{2^m} (v\in V_c)$ sent from its $k$ adjacent variable nodes. 
The check node $c$ sends the message $P^{(\ell+1)}_{cv}\in [-1,1]^{2^m}$ to $v$ for $v\in V_c$.
\begin{align*}
&\tilde{P}^{(\ell)}_{vc}(z) = {P}^{(\ell)}_{vc}(H_{cv} z), \text{ for } z\in \GF(2^m)\nonumber\\
&\tilde{P}^{(\ell+1)}_{cv}(z) = \prod_{v'\in V_c\backslash\{v\}}\tilde{P}^{(\ell)}_{v'c}(z) \text{ for } z\in \GF(2^m),\\
&{P}^{(\ell+1)}_{cv}(z) = \tilde{P}^{(\ell+1)}_{cv}(H_{cv}^{-1} z),\text{ for } x\in \GF(2^m)\nonumber.
 \end{align*} 
In \ref{102047_13Feb10}, we give 
the definition of $H_{cv}$ and the explanation that this step is equivalent to the check-to-variable step of the SP algorithm.  
Increment the iteration round as $\ell:=\ell+1$. 
\\\\\vtoc:
Each variable node $v$ has $j$ incoming messages $P^{(\ell)}_{cv} (c\in C_v)\in [-1,1]^{2^m}$ sent from its $j$ adjacent check nodes. 
The variable node $v$ sends the message $p^{(\ell)}_{vc}\in [-1,1]^{2^m}$ to $c$ for $c\in C_v$.
\begin{align*}
&\tilde{P}^{(\ell+1)}_{cv}= P_v^{(0)}\bigotimes_{c'\in C_v\backslash\{c\}}\tilde{P}^{(\ell)}_{c'v},
\end{align*}
where $P_1\otimes P_2\in [-1,1]^{2^m}$ is a convolution of $P_1\in [-1,1]^{2^m}$ and $P_2\in [-1,1]^{2^m}$. To be precise,  for  $x\in \GF(2^m)$
\begin{equation*}
  (P_1\otimes P_2)(z) = \sum_{z_1,z_2\in\GF(2^m):z=z_1+z_2}{P_1(z_1)P_2(z_2)}.
\end{equation*}
And normalize $P^{(\ell)}_{vc}\in [-1,1]^{2^m}$ so that $P^{(\ell)}_{vc}(0)=1$ as follows. 
\begin{align*}
 P^{(\ell)}_{vc}(z):={P^{(\ell)}_{vc}}(z)/{P^{(\ell)}_{vc}(0)}, \text{ for } z\in\GF(2^m). 
\end{align*}
\tentative:
The tentatively estimated symbol $\hat{x}_v^{(\ell)}$  for the $v$-th transmitted symbol is given as 
\begin{align}
\hat{x}_v^{(\ell)}:=\mathop{\mathrm{argmax}}_{x\in \GF(2^m)}q^{(\ell)}(x),\label{141405_13Feb10} 
\end{align} 
where for $x, z\in \GF(2^m),$
\begin{align}
&Q_v^{(\ell)}(z)= P_v^{(0)}(z)\prod_{c'\in C_v}P^{(\ell)}_{c'v}(z),\nonumber\\
&q_v^{(\ell)}(x)= \sum_{x\in\GF(2^m)}Q_v^{(\ell)}(z)(-1)^{z\cdot x}.\label{141552_13Feb10}
\end{align}

It is cumbersome that we have to apply the inverse Fourier transform $Q_v^{(\ell)}$ as in Eq.~\eqref{141552_13Feb10} to decide tentatively estimated symbols in Eq.~\eqref{141405_13Feb10}.
We give an alternative way of determining estimated symbols, which does not need the inverse Fourier transform. 

It is known \cite{mct} that Eq.~\eqref{141405_13Feb10} gives the MAP symbol for the $v$-th transmitted symbol, when the Tanner graph is a tree.  
When the Tanner graph is not a tree, the approximated MAP symbol is obtained due to the sparseness of the Tanner graph. 
The symbol-MAP decoder minimizes the symbol error rate (SER) while the bit-MAP decoder minimizes the bit error rate (BER). 
For the digital communications, it is widely desirable to lower the BER rather than the SER. 
For $x=(x_1,\dotsc, x_m)\in\GF(2^m)$, by marginalizing $q_v^{(\ell)}$ as
\begin{align*}
&q_{v,i}^{(\ell)}(x_i):=\sum_{x_1,\dotsc,x_{i-1},x_{i+1},\dotsc, x_m\in\GF(2)}q_v^{(\ell)}(x), 
\end{align*} 
the approximated MAP bit $\hat{x}_{v,i}\in\GF(2)$  for the $i$-th bit in the $v$-th transmitted symbol is obtained  by
\begin{align*}
&\hat{x}^{(\ell)}_{v,i}:=\mathop{\mathrm{argmax}}_{x_i\in \GF(2)}q_{v,i}^{(\ell)}(x_i).
\end{align*} 
Let $\alpha$ be the fixed primitive element of $\GF(2^m)$. 
For $i=1,\dotsc,m$,  $\alpha^{i-1}\in\GF(2^m)$ is represented as a $m$-bit sequence
$(\overbrace{0,\dotsc,0}^{i-1},1,\overbrace{0,\dotsc,0}^{m-i})$. 
It follows that 
\begin{align*}
 Q_v^{(\ell)}(\alpha^{i-1})&=\sum_{x\in\GF(2^m)}q_v^{(\ell)}(x)(-1)^{\alpha^{i-1}\cdot x}\\
&=\sum_{x\in\GF(2^m)}q_v^{(\ell)}(x)(-1)^{x_i}\\
&=q_{v,i}^{(\ell)}(0)-q_{v,i}^{(\ell)}(1).
\end{align*}
Thus, 
without the inverse Fourier transform, directly from $Q_v^{(\ell)}\in [-1,1]^{2^m}$, 
we can calculate
the approximated MAP bit $\hat{x}_{v,i}$ as
\begin{align}
\label{mapbitfsp}
 &\hat{x}_{v,i}^{(\ell)}=
\left\{\begin{array}{ll}
   0 & \text{if }Q_v^{(\ell)}(\alpha^{i-1})>0,\\
   1 & \text{if }Q_v^{(\ell)}(\alpha^{i-1})<0.
\end{array}
\right.
\end{align}

In a similar way, we can calculate the syndromes without the inverse Fourier transform. 
For given estimated symbol sequence $(\hat{x}_1^{(\ell)},\dotsc, \hat{x}_N^{(\ell)})\in\GF(2^m)^N$ with $\hat{x}_v^{(\ell)}=(\hat{x}_{v,1}^{(\ell)},\dotsc,\hat{x}_{v,m}^{(\ell)})\in\GF(2)^{mN}$, 
the syndrome symbol for a check node $c$ is given by 
\begin{align*}
\hat{s}_c^{(\ell)}:=\sum_{v\in V_c}h_{cv}\hat{x}_v^{(\ell)}. 
\end{align*}
In a similar way, it can be shown that $i$-th bit $\hat{s}_{c,i}$ of the syndrome symbol $\hat{s}_c$
is given as 
\begin{align*}
 &\hat{s}_{c,i}^{(\ell)}=
\left\{\begin{array}{ll}
   0 & \text{if }\prod_{v\in V_c}Q_v^{(\ell)}(H_{cv}\alpha^{i-1})>0\\
   1 & \text{if }\prod_{v\in V_c}Q_v^{(\ell)}(H_{cv}\alpha^{i-1})<0, 
\end{array}
\right.
\end{align*}
The sequence of the estimated symbols $(\hat{x}_{1}^{(\ell)},\dotsc, \hat{x}_{v}^{(\ell)})\in\GF(2^m)^N$ forms a codeword if 
$\hat{s}_{c,i}^{(\ell)}=0$ for all $c=1,\dotsc,M$ and $i=1,\dotsc, m$. 
\subsection{New Log-Fourier Sum-Product Algorithm for Non-Binary LDPC Codes}
\label{logsp}
In this section, we proposed  the Fourier SP algorithm operated in the logarithm domain. 
The multiplications in the Fourier SP algorithm are replaced with additions in the in the logarithm domain. 

The Fourier SP algorithm requires many multiplications which can not be affordable for 
realizing the high speed and low quantization level  decoders.  
In an analogous way as in Log-SP algorithm, we can consider the Fourier-SP algorithm in the logarithm domain, which is referred to as Log-Fourier SP algorithm. 

For the sake of simple description of the algorithm, and in order to emphasize the analogy with 
the Fourier SP algorithm, 
we use a logarithm-like function $\Gamma:[-1,1]\to \GF(2)\times[-\infty,0]$ as follows. 
\begin{align*}
  &\Gamma(x):= (\sgn_{\GF(2)}(x), \ln(|x|))\in \GF(2)\times[-\infty,0],\\
 &\sgn_{\GF(2)}(x):=\left\{
 \begin{array}{ll}
  0\in\GF(2) & (x>0)\\
  \mbox{choose randomly } 0 \mbox{ or }   1 & (x=0)\\
  1\in\GF(2) & (x<0).
 \end{array}\right.
\end{align*}
Obviously, for any non-zero real numbers $x,y$ it holds that $\Gamma(xy)=\Gamma(x)+\Gamma(y)$ and
$\Gamma^{-1}(\cdot)$ is well-defined. 
The following describes the proposed Log-Fourier domain decoding of non-binary LDPC codes. 
\\\\\initialization:
For each variable node $v$, the initial message $\Lambda_v^{(0)}\in (\GF(2)\times[-\infty,0])^{2^m}$ is given as follows. 
 \begin{align*}
  p_v^{(0)}(x)&= \Pr(X_v=x|Y_v=y_v), x\in\GF(2^m)\\
  P_v^{(0)}(z)&= \sum_{x\in\GF(2^m)} {p_v^{(0)}(z)}(-1)^{z\cdot x}, z\in\GF(2^m),\\
  \Lambda_v^{(0)}(z)&= \Gamma (P_v^{(0)}(z)), z\in\GF(2^m).
 \end{align*} 
Each variable node $v$ sends the message $\Lambda^{(0)}_{vc}=\Lambda^{(0)}_{v}\in (\GF(2)\times[-\infty,0])^{2^m}$ to $c$ for $c\in C_v$.
Set $\ell=0$.
\\\\\ctov:
Each check node $c$ has $k$ incoming messages $\Lambda^{(\ell)}_{vc}\in (\GF(2)\times[-\infty,0])^{2^m} (v\in V_c)$ sent from its $k$ adjacent variable nodes. 
The check node $c$ sends the message $\Lambda^{(\ell+1)}_{cv}\in (\GF(2)\times[-\infty,0])^{2^m}$ to $v$ for $v\in V_c$.
\begin{align*}
&\tilde{\Lambda}^{(\ell)}_{vc}(z) = {\Lambda}^{(\ell)}_{vc}(H_{cv}z),\\
&\tilde{\Lambda}^{(\ell+1)}_{cv} = \sum_{v'\in V_c\backslash\{v\}}\tilde{\Lambda}^{(\ell)}_{v'c}(z) \text{ for } z\in \GF(2^m)\\
&{\Lambda}^{(\ell+1)}_{cv}(z) = \tilde{\Lambda}^{(\ell+1)}_{cv}(H_{cv}^{-1}z) .
\end{align*} 
Increment the iteration round as $\ell:=\ell+1$. 
\\\\\vtoc:
Each variable node $v$ has $j$ incoming messages $\Lambda^{(\ell)}_{cv} (c\in C_v)\in (\GF(2)\times[-\infty,0])^{2^m}$ sent from its $j$ adjacent check nodes. 
The variable node $v$ sends the message $\Lambda^{(\ell)}_{vc}\in (\GF(2)\times[-\infty,0])^{2^m}$ to $c$ for $c\in C_v$.
\begin{align*}
\Lambda_v^{(\ell)}(z)= \Lambda_v^{(0)}(z){\Huge \boxplus}_{c'\in V_c\backslash\{c\}}\Lambda^{(\ell)}_{c'v}(z), 
 \end{align*} 
where $\Lambda_1\boxplus \Lambda_2\in (\GF(2)\times[-\infty,0])^{2^m}$ is defined as follows. 
\begin{align}
 &(\lambda_1\boxplus \lambda_2)(x)\label{230321_22May10} \\
&=\Gamma( \sum_{x_1,x_2\in\GF(2^m):x=x_1+x_2}\Gamma^{-1}({\lambda_1(x_1)+\lambda_2(x_2)}))\nonumber,
\end{align} 
for $x\in\GF(2^m)$.
The difference between the 2 operators $\boxtimes$ and $\boxplus$ is only a sign bit, 
which can be ignored.  
We also refer to the operator $\cdot\boxplus\cdot$ as the log-convolution.
\\\\\tentative:
The tentatively estimated symbol $\hat{x}_v^{(\ell)}$ for the $v$-th symbol is given as 
\begin{align*}
&\hat{x}_v^{(\ell)}=\mathop{\mathrm{argmax}}_{x\in \GF(2^m)}\mu_v^{(\ell)}(x)\\
& \mu_v^{(\ell)}(x):= \sum_{z\in\GF(2^m)}M_v^{(\ell)}(z)(-1)^{z\cdot x}\\
& M_v^{(\ell)}(z):= \Lambda_v^{(0)}(z){\Huge \boxplus}_{c'\in V_c}\Lambda^{(\ell)}_{c'v}(z).
\end{align*}

We calculated the MAP bit for the Fourier SP algorithm in Eq.~\eqref{mapbitfsp}. 
In a similar way, we can calculate
the MAP bit for the $i$-th bit in the $v$-th transmitted symbol $\hat{x}_{v,i}^{(\ell)}$, 
without the inverse Fourier transform, directly from $M_v^{(\ell)}\in (\GF(2)\times [-\infty,0])^{2^m}$ as
\begin{align*}
 &\hat{x}_{v,i}^{(\ell)}= \text{the first entry of } M_v^{(\ell)}(\alpha^{i-1}).
\end{align*}

In a similar way, 
it can be shown that $i$-th bit $\hat{s}_{c,i}^{(\ell)}\in \GF(2)$ of the syndrome symbol $\hat{s}_{c}^{(\ell)}\in\GF(2^m)$ of a check node $c$
for the estimated MAP bits $\hat{x}_{v,i}^{(\ell)}$ ($v=1,\dotsc,N, i=1,\dotsc,m$)
is calculated, 
without the inverse Fourier transform as
\begin{align*}
 &\hat{s}_{c,i}^{(\ell)}=
\text{the first entry of } \sum_{v\in V_c}M_v^{(\ell)}(H_{cv}\alpha^{i-1}),
\end{align*}
for $c=1,\dotsc,M$ and $i=1,\dotsc,m$. 
The sequence of the estimated symbols $(\hat{x}_{1}^{(\ell)},\dotsc, \hat{x}_{v}^{(\ell)})$ forms a codeword if 
$\hat{s}_{c,i}=0$ for all $c=1,\dotsc,M$ and $i=1,\dotsc, m$. 
\section{Comparison of Computation Amount}
In this section, we compare the computation amount of the conventional and proposed algorithms. 

In the conventional Log-SP algorithm, 
for each check node $c$, with the $k$ incoming  messages $\lambda^{(\ell)}_{vc}\in [-\infty,0]^{2^m}$ for $v\in V_c$, 
$c$ needs to compute a  $(k-1)$-fold log-convolution
\begin{align*}
 \boxtimes_{v'\in V_c\backslash\{v\}}\tilde{\lambda}^{(\ell)}_{v'c}
\end{align*}
for $v\in V_c$, i.e., for $k$ times. 
For each variable node $v$, with the $j$ incoming  messages 
$\lambda^{(\ell)}_{cv}\in [-\infty,0]^{2^m}$ for $c\in C_v$, 
$v$ needs to compute a  $j$-term component-wise addition
\begin{align*}
 \lambda_v^{(0)}(x)+\sum_{c'\in C_v\backslash\{c\}}\lambda^{(\ell)}_{c'v}(x) \text{ for } x\in \GF(2^m). 
\end{align*}
for $v\in V_c$, i.e., for $j$ times. 

While, in the proposed Log-Fourier SP algorithm, 
for each check node $c$, with the $k$  incoming  messages $\Lambda^{(\ell)}_{vc}\in (\GF(2),[-\infty,0])^{2^m}$ for $v\in V_c$, 
$c$ needs to compute a $(k-1)$-term component-wise addition
\begin{align*}
 \sum_{v'\in V_c\backslash\{v\}}\tilde{\lambda}^{(\ell)}_{v'c}(z) \text{ for } z\in \GF(2^m),
\end{align*}
for $v\in V_c$, i.e., for $k$ times. 
For each variable node $v$, with the $j$ incoming  messages 
$\Lambda^{(\ell)}_{cv}\in [-\infty,0]^{2^m}$ for $c\in C_v$, 
$v$ needs to compute a $j$-fold log-convolution
\begin{align*}
 \Lambda_v^{(0)}{\Huge \boxplus}_{c'\in C_v\backslash\{c\}}\Lambda^{(\ell)}_{c'v}
\end{align*}
for $v\in V_c$, i.e., for $j$ times. 

The aim of this paper is reducing the most complex node-computation among all the node for the node-parallel implementation. 
And the most complex node-computation in both  conventional Log-SP and proposed Log-Fourier SP is the log-convolution.
Indeed, one component-wise addition of vector of length $2^m$ requires only $2^{m}$ additions. 
On the other hand, one log-convolution requires as much as $2^{2m}$ computations of additions, $\ln(\cdot)$ and $\exp(\cdot)$. 
Consequently, we focus our attention to the computation amount of log-convolutions for both algorithms. 
We assume we use a $(j=2,k)$-regular non-binary LDPC code over $\GF(2^m)$, 
since it is empirically known that good non-binary LDPC codes have  parity-check matrices of column weight 2 \cite{4641893}. 
This property is extremely preferable for the Log-Fourier SP algorithm, 
since it is only needs 2 times log-convolutions per variable node. 
Table \ref{171815_14Feb10}  compares
the computation amount per node  for $(2,k)$-regular non-binary LDPC codes over $\GF(2^m)$ for the conventional Log-SP and the proposed Log-Fourier SP algorithm. 
It can be seen that the proposed Log-Fourier SP algorithm needs only a constant number of the log-convolutions per node even if $k$ gets lager to increase the coding rate $R=(k-2)/k$. 
It can be seen that the proposed Log-Fourier SP algorithm needs less log-convolutions per node. 

Due to the intensive computation and the large number of incoming messages, 
the log-convolutions have been 
the main obstacles blocking parallelized implementations of high speed decoders for LDPC-codes.  
The number of necessary component-wise additions per node in the proposed Log-Fourier SP algorithm is larger that in the conventional Log-SP algorithm. 
Nevertheless, the computation amount of log-convolutions per node is largely reduced. 
With the Log-Fourier SP algorithm, we can realize the node-parallel implementation which does not have the bottleneck problem. 

In the situation that each node-computation is processed in parallel, 
the total decoding time depends on the most complex computation among the all nodes. 
The proposed Log-Fourier SP algorithm 
can reduce the largest computation amount of among all the nodes. 
To be  precise, the $(k-1)$-fold log-convolutions for $k$ times were the most 
complex node-computation in the conventional Lot-SP algorithm. 
The most complex node-computation in the Log-Fourier SP algorithm 
is reduced to $2$-fold log-convolutions for 2 times.
 \begin{table*}[t]
    \capwidth=0.7\textwidth
\begin{center}
 \caption{Comparison of the computation amount per node  for $(2,k)$-regular non-binary LDPC codes over $\GF(2^m)$. 
ADD stands for the component-wise addition of vectors of length $2^m$. 
CONV stands for the log-convolution of vectors of length $2^m$, as defined in Eq.~\eqref{230321_22May10}. 
Usually, $k\ge 3$ is used. 
One ADD requires only $2^{m}$ additions. 
On the other hand, one CONV requires as much as 
$2^{2m}$ computations of  additions, $\ln(\cdot)$ and $\exp(\cdot)$. 
}
 \label{171815_14Feb10}
  \begin{tabular}{|l|l|l|l|}
  \hline
                & \vtoc   &  \ctov    & \tentative \\\hline
 Log-SP         & $2$-term ADD $\times 2$          &  $(k-1)$-fold \bf{CONV}$\times k$ &$3$-term ADD\\\hline
 Log-Fourier SP & $2$-fold \bf{CONV} $\times 2$         &  $(k-1)$-term ADD$\times k$ &$3$-fold \bf{CONV}\\\hline
 \end{tabular}
\end{center}
 \end{table*}

\section{Discussions and Conclusions}
In this paper, we proposed a decoding algorithm   suitable for fully node-parallel implementation of non-binary LDPC codes.
The proposed algorithm reduces the most complex node-computation, which results large reduction of 
the total decoding time in the situation that each node-computation is processed in parallel. 


It should be noted that Hartman and Rudolph (HR) \cite{1055617} developed the decoding algorithm for the dual code by using Fourier-transform.
The HR decoding makes the MAP decoding  possible by decoding the dual codes with the Fourier transformed channel outputs.
However, the application of HR decoding to LDPC codes have been  limited to the decoding the constituent high-rate codes, e.g.~single parity-check codes for LDPC codes. 
Gallager's $f$ function \cite[pp.~43]{gallager_LDPC} can be viewed as Fourier transforming log likelihood ratio (LLR) to the Log-Fourier domain, which 
reduce decoding of a single parity-check code to decoding a repetition code. 
Isaka used the HR decoding the constituent  Hamming codes for the generalized LDPC codes \cite{1388739}. 
The dual code of LDPC code with a factor graph $G$ is given by replacing 
``='' nodes and ``+'' nodes \cite{910573}.
Dual of an LDPC code with parity-check matrix $H$ 
is given by a low-density generator-matrix (LDGM) code with parity-check matrix $(H^T|I)$ by puncturing the bits corresponding to $H^T$. 
The proposed algorithm can be viewed as a slightly modified application of the HR decoding, not to the constituent codes, but to the whole LDPC code.  
The modification is that the decoding algorithm for the dual code of the LDPC code is not the MAP decoding but the Log-SP algorithm.



\label{054009_6Oct09}
\appendix
\section{Companion Matrix}
\label{102047_13Feb10}
For the primitive elements $\alpha\in\GF(2^m)$, 
we denote the corresponding primitive polynomial by $\pi(x)=\pi_0+\pi_1x+\cdots+\pi_{m-1}x^{m-1}+x^m$, 
where $\pi_0,\dotsc,\pi_{m-1}\in\GF(2)$.
The companion matrix of $\alpha$ is given as
\begin{align*}
     A=\begin{bmatrix} 0 & 0 & \dots & 0 & \pi_0 \\ 1 & 0 & \dots & 0 & \pi_1 \\ 0 & 1 & \dots & 0 & \pi_2 \\ \vdots & \vdots & \vdots & \vdots & \vdots \\ 0 & 0 & \dots & 1 & \pi_{m-1} \\ 
\end{bmatrix}. 
\end{align*}
As shown in \cite{macwilliams77}, under $m$-bit representation of $\GF(2^m)$ symbols, it is readily checked that 
\begin{align*}
 A^i\alpha^j=\alpha^{i+j}=\alpha^{i}\alpha^{j},
\end{align*}
where the $\alpha^j$ and $\alpha^{i+j}$ are interpreted as a $m$-bit vectors. 
For $h_{cv}=\alpha^i$, we define $H_{cv}$ as a $m\times m$ binary matrix  $H_{cv}=(A^i)^T$.
Then we have $h_{cv}x=H_{cv}x$ and $h_{cv}^{-1}x=H_{cv}^{-1}x$. 
We will show the check-to-variable step of the Fourier SP algorithm is equivalent with that of the SP algorithm. 
To this end, it is sufficient to show that the Fourier transform of $\tilde{p}^{(\ell)}_{vc}$ is $\tilde{P}^{(\ell)}_{vc}$. 
\begin{align*}
&  \sum_{x\in\GF(2^m)}\tilde{p}^{(\ell)}_{vc}(x)(-1)^{z\cdot x}=\sum_{x\in\GF(2^m)}{p}^{(\ell)}_{vc}(h_{cv}^{-1}x)(-1)^{z\cdot x}\\
&=\sum_{x\in\GF(2^m)}{p}^{(\ell)}_{vc}(x)(-1)^{z\cdot (H_{cv}x)}\\
&=\sum_{x\in\GF(2^m)}{p}^{(\ell)}_{vc}(x)(-1)^{(H_{cv}^Tz)\cdot x}=P(H_{cv}^Tz)=\tilde{P}(z). 
\end{align*}

\bibliographystyle{ieicetr}
\bibliography{IEEEabrv,kenta_bib}
\end{document}